\documentclass[a4paper,onecolumn,notitlepage,preprintnumbers,amsmath,amssymb,amsfonts,floatfix,nofootinbib,showkeys]{revtex4-1} 

\usepackage{dcolumn}
\usepackage{bm}
\usepackage{float}
\usepackage{epsfig}
\usepackage{graphicx}
\usepackage{longtable} 
\usepackage{rotating}
\usepackage{amsmath}
\usepackage{amsfonts}
\usepackage{amssymb}
\usepackage{xspace}
\usepackage{multirow}
\usepackage{mathrsfs}
\usepackage{calrsfs}
\usepackage{txfonts}
\usepackage{graphicx}
\usepackage{hypernat}
\usepackage{color}
\usepackage[utf8]{inputenc}
\usepackage[pdftex,colorlinks,linkcolor=red,citecolor=red,anchorcolor=red,urlcolor=red]{hyperref}
\usepackage{enumitem}

\providecommand{\gaga}{\gamma\,\gamma}

\providecommand{\madgraph}{{\sc madgraph}}
\providecommand{\pythia}{{\sc pythia}}

\providecommand{\fastjet}{{\sc fastjet}}

\newcommand{\sqrts}{\sqrt{s}}
\newcommand{\sqrtsnn}{\sqrt{s_{_{\textsc{nn}}}}}
\newcommand{\sqrtsgg}{\sqrt{s_{\gaga}^{\rm max}}}
\newcommand{\bbbar}    {\rm {b\bar{b}}}
\newcommand{\ccbar}    {\rm {c\bar{c}}}
\newcommand{\qqbar}    {\rm {q\bar{q}}}
\newcommand{\epem}{e^+e^-}
\newcommand{\cm}{c.m.\@\xspace}
\newcommand{\elm}{e.m.\@\xspace}
\newcommand{\ie}{i.e.\@\xspace}
\newcommand{\LumiInt}{\mathcal{L}_{\rm \tiny{int}}}

\begin{document}

\title{Higgs boson production in photon-photon interactions with proton, \\
light-ion, and heavy-ion beams at current and future colliders}

\date{\today}

\author{David d'Enterria\footnote{email: david.d'enterria@cern.ch}}
\affiliation{CERN, EP Department, 1211 Geneva, Switzerland}
\author{Daniel E. Martins\footnote{email: dan.ernani@gmail.com}}
\affiliation{UFRJ, Univ. Federal do Rio de Janeiro, 21941-901, Rio de Janeiro, RJ}
\author{Patricia Rebello Teles\footnote{email: patricia.rebello.teles@cern.ch}}
\affiliation{UERJ, Univ. do Estado do Rio de Janeiro, 20550-900, Rio de Janeiro, RJ}

\begin{abstract}\noindent
The production of the Higgs boson in photon-photon interactions with proton and nucleus 
beams at three planned or proposed future CERN colliders --- the high-luminosity Large Hadron Collider (HL-LHC), 
the high-energy LHC (HE-LHC), and the Future Circular Collider (FCC) --- is studied. 
The cross sections for the process AA$\xrightarrow{\gamma\gamma}$(A)\,H\,(A), with the ions A 
surviving the interaction and the Higgs scalar exclusively produced, are computed with \madgraph~5 
modified to include the corresponding elastic $\gamma$ fluxes, for Pb-Pb, Xe-Xe, Kr-Kr, Ar-Ar, O-O, p-Pb, 
and p-p over the nucleon-nucleon collision energy range $\sqrtsnn\approx 3$--100~TeV. 
Simulations of the $\gaga\to\,$H$\,\to\bbbar$ decay mode --- including realistic (mis)tagging and 
reconstruction efficiencies for the final-state b-jets, as well as appropriate kinematical selection criteria 
to reduce the similarly computed $\gaga\to\bbbar,\ccbar,\qqbar$ continuum backgrounds --- have been carried out. 
Taking into account the expected luminosities for all systems, the yields and significances 
for observing the Higgs boson in ultraperipheral collisions (UPCs) are estimated.
At the HL-LHC and HE-LHC, the colliding systems with larger Higgs significance are Ar-Ar(6.3~TeV) and Kr-Kr(12.5~TeV)
respectively, but $3\sigma$ evidence for two-photon Higgs production would require 200 and 30 times larger 
integrated luminosities than those planned today at both machines. Factors of ten can be gained by running for a year, 
rather than the typical 1-month heavy-ion LHC operation, but the process will likely remain unobserved until 
a higher energy hadron collider, such as the FCC, is built. In the latter machine, the $5\sigma$ observation of Higgs production 
in UPCs is feasible in just the first nominal run of Pb-Pb and p-Pb collisions at $\sqrtsnn = 39$ and 63~TeV respectively. 
\end{abstract}

\pacs{14.80.Bn, 25.20.Lj}

\maketitle


\section{Introduction}

Heavy ions accelerated at high energies are surrounded by huge electromagnetic (\elm) fields generated by the collective action 
of their $Z$ individual proton charges. In the equivalent photon approximation (EPA)~\cite{WW}, such strong \elm\
fields can be identified as quasireal photon beams with very low virtualities $Q^{2} < 1/R_A^{2}$ and large longitudinal 
energies of up to $\omega_{\rm max}\approx\gamma_L/R_A$, where $R_A$ is the radius of the charge and $\gamma_L=E_{\rm beam}/m_{N,p}$ 
is the beam Lorentz factor for nucleon or proton mass $m_{N,p} = 0.9315,\,0.9382$~GeV~\cite{Bertulani:1987tz,Baltz:2007kq}. 
On the one hand, since the photon flux scales as the squared charge of each colliding particle, 
photon-photon cross sections are enhanced millions of times for heavy ions (up to $Z^4 \approx 5\cdot 10^{7}$ for Pb-Pb) compared to proton or 
electron beams. On the other, proton (and lighter ions) feature larger $\omega_{\rm max}$ values thanks to their lower radii $R_A$ and larger
beam $\gamma_L$ factors, and can thereby reach higher photon-photon center-of-mass (\cm) energies. At the energies of the Large Hadron Collider (LHC), 
photons emitted from nuclei (with radii $R_A\approx 1.2\,A^{1/3}$~fm) are almost on-shell (virtuality $Q<$~0.06~GeV, for mass numbers $A>$~16), 
and reach longitudinal energies of up to hundreds of GeV, whereas photon fluxes from protons ($R_A\approx$~0.7~fm) have larger virtualities, 
$Q\approx$~0.28~GeV, and longitudinal energies in the TeV range~\cite{Baltz:2007kq}. Table~\ref{tab:1} summarizes the 
relevant characteristics of photon-photon collisions in ultraperipheral collisions (UPCs) of proton and nuclear beams at 
three planned or proposed CERN future hadron colliders: the high-luminosity LHC (HL-LHC)~\cite{HL_LHC_HE_LHC_AA}, 
the high-energy LHC (HE-LHC)~\cite{HL_LHC_HE_LHC_AA,HE_LHC}, and the Future Circular Collider (FCC)~\cite{FCC_AA}. 
The beam luminosities for light- and heavy-ions considered here are those discussed in Refs.~\cite{HL_LHC_HE_LHC_AA,FCC_AA}.
Although the beam luminosities for p-p are 7 orders of magnitude larger than those for Pb-Pb, the running conditions with 
multiple pileup p-p collisions per bunch crossing hinder the measurement of exclusive $\gaga$ interactions with central masses at 
125-GeV (unless one installs, in the LHC case, very forward proton taggers at 420~m inside the tunnel\footnote{We note that a 
similar forward tagging of lead ions at the LHC is impossible, given that the ions carry a much larger longitudinal momentum,
$p_L$~=~5.5~TeV$\times$A $\approx$~600 TeV, than the protons, and thereby are barely deflected after a photon-photon interaction.}, 
with 10-picosecond time resolution~\cite{Albrow:2008pn}). 
Thus, in the present study we take $\LumiInt$~=~1~fb$^{-1}$ as the value potentially integrated 
under low-pileup conditions that allow the reconstruction of exclusive photon-photon final states in p-p collisions. 
In all cases in Table~\ref{tab:1}, one can see that the maximum photon-photon \cm\ energy reaches above the kinematical 
threshold for Higgs boson production, $\sqrtsgg\gtrsim m_{\rm H} = 125$\,GeV, through the process depicted in Fig.~\ref{tab:1} (left).
The observation of the $\gaga\to$~H process would provide, first, an independent measurement of the H--photon loop-induced 
coupling based not on the Higgs decay (as measured at the LHC~\cite{Higgs_observ}) but on its $s$-channel production mode. 
In addition, precise measurements of the $\rm \Gamma(H\to\gaga)$ partial width derived from 
$\rm \sigma(\gaga\to H\to\bbbar)\propto\Gamma(H\to\gaga)\cdot BR(H\to\bbbar)$, and of the Higgs branching 
ratio BR(H$\to\gaga$) determined at a future $\epem$ collider, would also provide a model-independent extraction of the 
total Higgs width, via $\rm\Gamma_{\rm H}^{\rm tot}=\Gamma(H\to\gaga)/BR(H\to\gaga)$~\cite{Borden:1993cw}.



\begin{figure}[H]
\centering
\includegraphics[width=0.5\columnwidth]{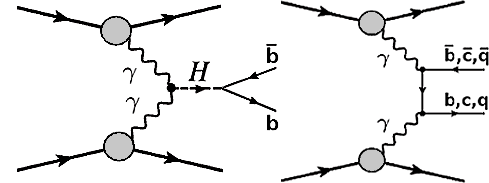}
\caption{\label{fig:1}Diagrams for the exclusive two-photon production of the Higgs boson (followed by its $\bbbar$ decay, left), 
and of b-,c-,light-quark pairs (processes that share the same final state as the Higgs case, right) in ultraperipheral proton/nuclear collisions.}
\end{figure}

\vspace*{-0.2cm}
\begin{table}[H]
\centering
\caption[]{Summary of the characteristics of photon-photon collisions in ultraperipheral proton and nuclear collisions at 
the HL-LHC, HE-LHC, and FCC~\cite{HL_LHC_HE_LHC_AA,HE_LHC,FCC_AA}: 
(i) Nucleon-nucleon \cm\ energy $\sqrtsnn$, (ii) integrated luminosity per run $\LumiInt$, 
(iii) beam energies $\rm E_{beam}$, (iv) Lorentz factor $\gamma_L$, (v) effective charge radius $R_A$, 
(vi) photon ``maximum'' energy $\omega_{\rm max}$ in the \cm\ frame, and
(vii) ``maximum'' photon-photon \cm\ energy $\sqrt{s_{\gaga}^{\rm max}}$.
The last two columns list the $\rm \gaga\to H$ cross sections and the 
expected number of Higgs events for the quoted $\LumiInt$ per system.}
\label{tab:1}
\vspace{0.2cm}
\resizebox{\textwidth}{!}{%
\begin{tabular}{lccccccccc} \hline
System\!\!  & \!\!$\sqrtsnn$\!\! & \!\!$\LumiInt$\!\! & \!\!\!\!$E_{\rm beam1}+E_{\rm beam2}$\!\! & $\gamma_L$ &  $R_A$ & $\omega_{\rm max}$ 
       & $\sqrt{s_{\gaga}^{\rm max}}$ & $\rm \sigma(\gamma\gamma\to H)$ & $\rm N(\gamma\gamma\to H)$ \\
\hline
Pb-Pb & 5.5  TeV &  10~nb$^{-1}$ & 2.75 + 2.75 TeV & 2950 & 7.1 fm &  80 GeV & 160 GeV & 15~pb & 0.15 \\ 
Xe-Xe & 5.86 TeV &  30~nb$^{-1}$ & 2.93 + 2.93 TeV & 3150 & 6.1 fm & 100 GeV & 200 GeV &  7~pb & 0.21 \\ 
Kr-Kr & 6.46 TeV & 120~nb$^{-1}$ & 3.23 + 3.23 TeV & 3470 & 5.1 fm & 136 GeV & 272 GeV &  3~pb & 0.36 \\ 
Ar-Ar & 6.3  TeV & 1.1~pb$^{-1}$ & 3.15 + 3.15 TeV & 3400 & 4.1 fm & 165 GeV & 330 GeV & 0.36~pb & 0.40 \\ 
O-O   & 7.0  TeV & 3.0~pb$^{-1}$ & 3.5 + 3.5 TeV   & 3750 & 3.1 fm & 240 GeV & 490 GeV & 35~fb & 0.11 \\ 
p-Pb  & 8.8 TeV  &   1~pb$^{-1}$ & 7.0 + 2.75 TeV  & 7450, 2950 & 0.7, 7.1 fm & 2.45 TeV, 130 GeV & 2.6 TeV & 0.17~pb & 0.17 \\ 
p-p   & 14 TeV   &   1~fb$^{-1}$ & 7.0 + 7.0 TeV   & 7450 & 0.7 fm & 2.45 TeV& 4.5 TeV & 0.18~fb & 0.18 \\ \hline 

Pb-Pb & 10.6 TeV &  10~nb$^{-1}$ & 5.3 + 5.3 TeV   & 5700 & 7.1 fm & 160 GeV & 320 GeV & 150~pb & 1.5 \\ 
Xe-Xe & 11.5 TeV &  30~nb$^{-1}$ & 5.75 + 5.75 TeV & 6200 & 6.1 fm & 200 GeV & 400 GeV &  60~pb & 1.8 \\ 
Kr-Kr & 12.5 TeV & 120~nb$^{-1}$ & 6.25 + 6.25 TeV & 6700 & 5.1 fm & 260 GeV & 530 GeV &  20~pb & 2.4 \\ 
Ar-Ar & 12.1 TeV & 1.1~pb$^{-1}$ & 6.05 + 6.05 TeV & 6500 & 4.1 fm & 320 GeV & 640 GeV & 1.7~pb & 1.9 \\ 
O-O   & 13.5 TeV & 3.0~pb$^{-1}$ & 6.75 + 6.75 TeV & 7300 & 3.1 fm & 470 GeV & 940 GeV & 0.11~pb & 0.33 \\ 
p-Pb  & 18.8 TeV &   1~pb$^{-1}$ & 13.5 + 5.3 TeV  & 14\,400, 5700 & 0.7, 7.1 fm & 4.1 TeV, 160 GeV & 4.2 TeV & 0.45~pb & 0.45 \\ 
p-p   & 27 TeV   &   1~fb$^{-1}$ & 13.5 + 13.5 TeV & 14\,400 & 0.7 fm & 4.1 TeV & 8.2 TeV & 0.30~fb & 0.30 \\ \hline 

Pb-Pb &  39 TeV  & 110~nb$^{-1}$ & 19.5 + 19.5 TeV & 21\,000 & 7.1 fm & 600 GeV & 1.2 TeV & 1.8~nb & 200 \\ 
p-Pb  &  63 TeV  &  29~pb$^{-1}$ & 50. + 19.5 TeV  & 53\,300, 21\,000 & 0.7,7.1 fm & 15.2 TeV, 600 GeV & 15.8 TeV &  1.5~pb & 45 \\ 
p-p   & 100 TeV  &   1~fb$^{-1}$ & 50. + 50. TeV   & 53\,300 & 0.7 fm & 15.2 TeV &  30.5 TeV & 0.70~fb & 0.70 \\\hline
\end{tabular}
}
\end{table}

The possibility to produce the Higgs boson by exploiting the huge photon fields in UPCs of ions,
AA$\xrightarrow{\gamma\gamma}$(A)H(A), where the scalar boson is produced at midrapidity and the colliding ions 
(A) survive their electromagnetic interaction (Fig.~\ref{fig:1} left), was first considered 30 years ago 
in several works~\cite{higgs_upc}. Detailed studies of the actual measurement of UPC-production of the 
Higgs boson in its dominant $\bbbar$ decay mode, including realistic experimental acceptance and efficiencies for the signal and 
the $\gaga\to\bbbar,\ccbar,\qqbar$ continuum backgrounds (Fig.~\ref{fig:1} right), were first presented in Ref.~\cite{dEnterria:2009cwl}
for ultraperipheral proton-nucleus (p-A) and nucleus-nucleus (A-A) collisions at LHC energies. This work showed that, for the nominal 
integrated luminosities, the scalar boson was unobservable in UPCs at the LHC unless one integrated at least 300 times more luminosity
than that expected for the standard 1-month heavy-ion operation. On the other hand, similar studies~\cite{dEnterria:2017qte} 
carried out within the FCC project, have indicated that the observation of Higgs production in UPCs was clearly 
possible in just the first nominal run of Pb-Pb and p-Pb collisions at $\sqrtsnn = 39$ and 63~TeV respectively. 
We note also that detailed studies of $\gaga\to$\,H were performed in the past in the context of the photon 
collider project~\cite{Borden:1993cw,PLC,NLO_bbbar}, exploiting the polarized and monochromatic $\gamma$ beams 
resulting from Compton-backscattering of laser light at future $\epem$ linear colliders, and with reduced backgrounds 
(thanks to the $\gamma$ polarization) compared to those considered in the present study. 
In this work here, we collect our UPC Higgs results carried out in the context of the FCC studies,
and discuss for the first time the conditions needed for an UPC Higgs boson measurement 
in the upcoming HL-LHC phase, as well as at the proposed HE-LHC with twice larger \cm\ energies.
Our new work includes not only higher luminosities than originally planned for the LHC, 
but also collisions of lighter ions (Xe-Xe, Kr-Kr, Ar-Ar, O-O) never considered before.

\section{Theoretical setup}
\label{sec:TH}

The \madgraph~5 (v.2.6.5) Monte Carlo (MC) event generator~\cite{madgraph} is employed to compute the UPC Higgs 
boson and diquark continuum cross sections, modified following the implementation discussed in detail in~\cite{dEnterria:2009cwl}, 
as well as to generate the corresponding events for subsequent analysis. The Higgs cross section is obtained from the convolution of the 
Weizs\"acker-Williams EPA photon fluxes for the proton and/or ions, and the elementary $\gaga \to \rm H$ cross section 
(with H-$\gamma$ coupling parametrized in the Higgs effective field theory~\cite{heft}), via
\begin{equation}
\sigma_{\rm A_1A_2 \to H} = \int dx_1\,dx_2\, f_{\gamma/A_1}(x_1) f_{\gamma/A_2}(x_2)\,\hat{\sigma}_{\gaga\to \rm H}\;,
\label{eq:sigma_UPC_H}
\end{equation}
where $x = \omega/E$ is the fraction of the energy of the incoming ion carried by each photon. 
The same expression is used for the $\bbbar,\ccbar,\qqbar$ continuums, where now the elementary 
$\hat{\sigma}_{\gaga \to \bbbar,\ccbar,\qqbar}$ cross sections at diquark invariant masses around
the Higgs mass are directly calculated at leading order (LO) by \madgraph~5. 
For protons, the \madgraph~5 default $\gamma$ flux is used, given by the energy spectrum of Ref.~\cite{Budnev:1974de}:
\begin{equation}
f_{\gamma/p}(x) = \frac{\alpha}{\pi} \, \frac{1 - x + 1/2 x^2}{x} 
\int_{Q_{\rm min}^2}^{\infty} \frac{Q^2 - Q_{\rm min}^2}{Q^4} | F(Q^2) |^2 dQ^2 \;,
\label{eq:f_x}
\end{equation}
where $\alpha=1/137$, $F(Q^2)$ is the proton \elm\ form factor, and the minimum momentum transfer 
$Q_{\rm min}$ is a function of $x$ and the proton mass $m_p$, $Q_{\rm min}^2 \approx (x\, m_p)^2/(1-x)$.
For ions of charge $Z$, the photon energy spectrum, integrated over impact parameter $b$ from $b_{\rm min}= R_A$ 
to infinity, is~\cite{Jackson}:
\begin{equation}
f_{\gamma/A}(x) = \frac{\alpha\,Z^2}{\pi} \, \frac{1}{x} \, \bigg[ 2 x_i K_0(x_i) K_1(x_i) - x_i^2 (K_1^2(x_i) - K_0^2(x_i)) \bigg] \; ,
\label{eq:flux_A}
\end{equation}
where $x_i= x\, m_N \, b_{\rm min}$, $K_0$, $K_1$ are the zero- and first-order modified Bessel functions of the second kind,
and for the different nuclear radii $R_A$, we use the data from elastic lepton-nucleus collisions~\cite{DeJager:1987qc}.
We exclude nuclear overlap by imposing $b_1 > R_{A_{1}}$ and $b_2 > R_{A_{2}}$ for each photon flux, and applying 
a correcting factor on the final cross section that depends on the ratio of Higgs mass over $\sqrtsnn$~\cite{Cahn:1990jk}. 

After cross section determination, the event generation is carried out for the dominant Higgs decay mode, 
$\rm H\to\bbbar$ with 56\% branching fraction~\cite{hdecay}, as it is the final state that provides the largest 
number of signal events. The same setup is used to generate the exclusive two-photon production of $\bbbar$ and (misidentified) 
$\ccbar$ and light-quark ($\qqbar$) jet pairs, which constitute the most important physical background for the 
H$\to \bbbar$ measurement. For the HL-LHC and HE-LHC systems, the analysis is carried out at the parton level only, 
whereas for FCC energies, we have further used \pythia~8.2~\cite{pythia8} to shower and hadronize the two final-state 
b-jets generated, which are then reconstructed with the Durham $k_{t}$ algorithm~\cite{kTalgo} (exclusive 2-jets final-state) 
using \fastjet~3.0~\cite{fastjet}. 
Given that the final state consists just of two quarks (jets) exclusively produced,
without any background that can potentially bias the four-momentum jet (quark) 
reconstruction, and that we take into account realistic jet resolution effects 
in the final dijet invariant mass analysis by appropriately smearing the parton-level 
results, no apparent differences exist between the parton- and hadron-level results,
as found previously in similar FCC-ee studies~\cite{Teles:2015xua,dEnterria:2017jmj} 
where, for the same set of kinematical cuts, both results are fully consistent within 
statistical uncertainties.

\section{Total Higgs cross sections}

\begin{figure}[H]
\centering
\includegraphics[width=0.47\columnwidth]{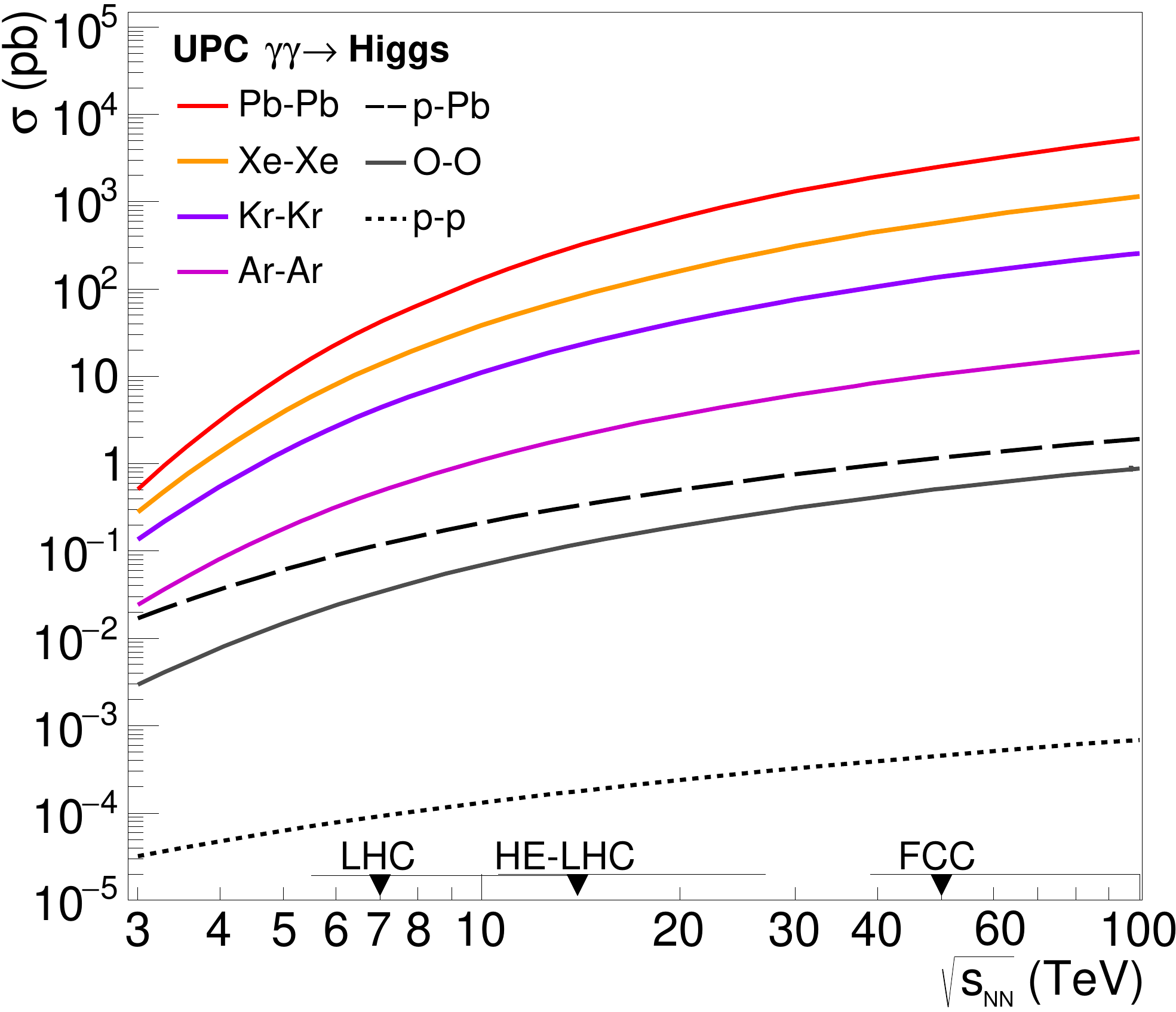}
\includegraphics[width=0.52\columnwidth]{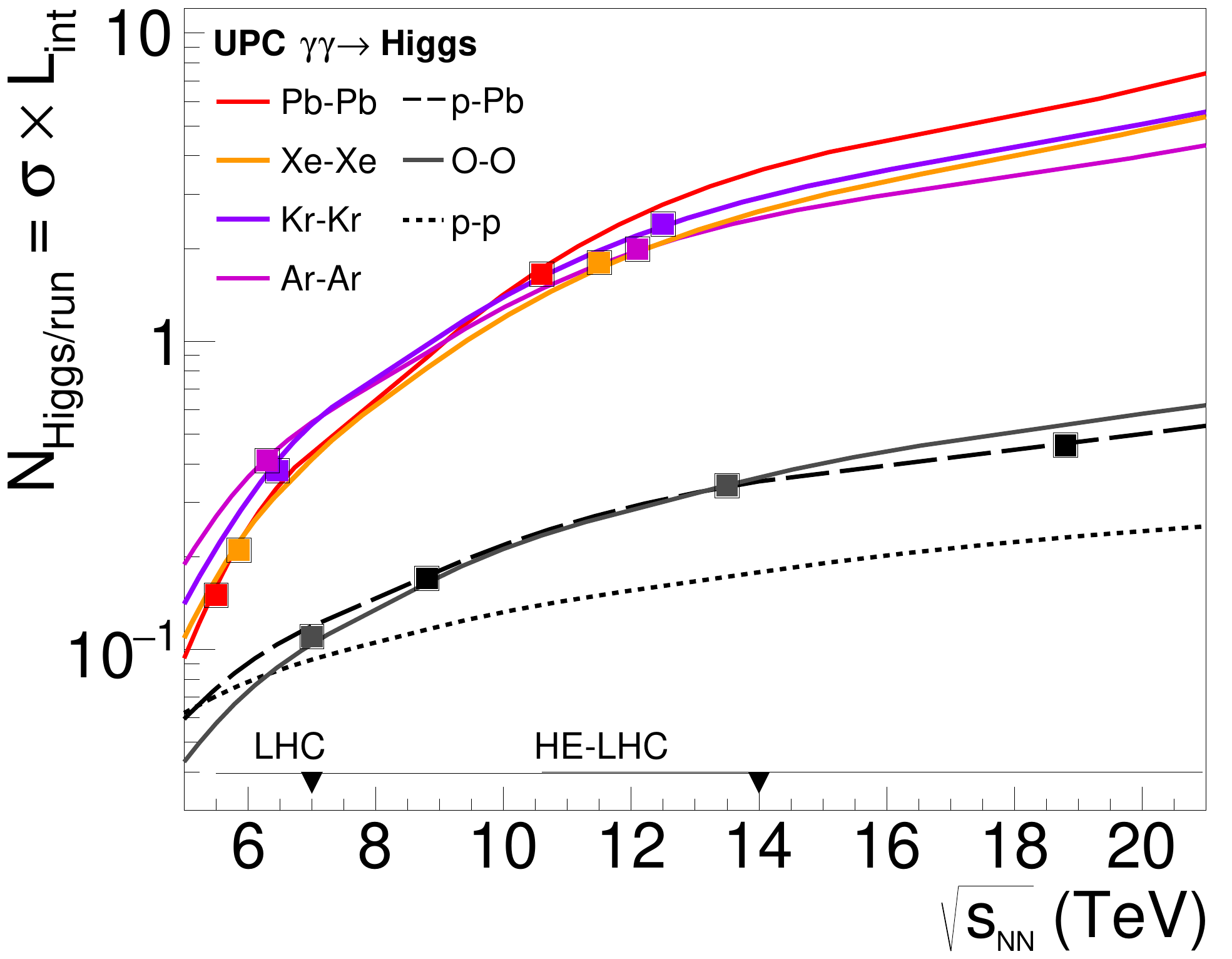}
\caption{\label{fig:2} Left: Two-photon fusion Higgs boson cross section versus nucleon-nucleon \cm\ energy 
in nuclear and proton collisions over $\sqrtsnn$~=~3--100~TeV. 
Right: Number of Higgs bosons produced per run (according to the integrated luminosities $\LumiInt$ listed in Table~\ref{tab:1}) 
in UPCs of various colliding systems in the HL-LHC and HE-LHC energy range. 
The square symbols indicate the nominal $\sqrtsnn$ for each colliding system.}
\end{figure}

The computed ultraperipheral Higgs boson cross sections as a function of $\sqrtsnn$ are shown in Fig.~\ref{fig:2} (left) 
and listed in the before-last column of Table~\ref{tab:1} for all p-p, p-A, and A-A systems considered. 
All theoretical cross sections have a conservative 20\% uncertainty (not quoted) to cover different charge form factors 
and nuclear overlap conditions~\cite{Klein:2016yzr,Harland-Lang:2018iur}. We note that the quoted cross sections are purely ``elastic'', 
\ie\ both incoming ions survive the \elm\ interaction. As discussed in~\cite{dEnterria:2009cwl}, $\gaga$ interactions can also be ``semielastic'',
and/or ``resolved'', with one (or both) quasireal photons being radiated from individual proton(s), and/or from individual quarks, inside the
colliding ions. In this latter case, one (or both) ions breakup at very forward rapidities after photon emission, 
AA$\xrightarrow{\gamma\gamma}$A\,H\,X, and the Higgs boson cross sections can be enhanced by about a
factor of two compared to the pure elastic results. We do not consider these cases here, and focus on the elastic processes
alone.\\

Figure~\ref{fig:2} (left) indicates that, as expected, the bigger the charge of the colliding ions, the larger the UPC 
Higgs cross sections, but such an advantage is mitigated in terms of final yields by the correspondingly reduced beam luminosities for heavier ions. 
Figure~\ref{fig:2} (right) shows the product of UPC Higgs cross section times the integrated luminosities for each
colliding system in the HL-LHC and HE-LHC energy range. At the LHC, we see that despite the fact that Pb-Pb features 
the largest Higgs cross section, $\rm \sigma(\gaga\to H)$~=~15~pb, there are about 2--3 times more scalar bosons produced
per month in Ar-Ar and Kr-Kr collisions (0.40 versus 0.15, last column of Table~\ref{tab:1}) thanks to the comparatively 
larger luminosities and \cm\ energies of the latter with respect to lead beams. 
At the HE-LHC, the Higgs cross sections are about a factor of 10 larger than at the LHC, and most colliding systems feature 
1.5--2.5 Higgs bosons produced per month. The most competitive systems to try a measurement of UPC Higgs production are
Ar-Ar at HL-LHC and Kr-Kr at HE-LHC respectively.
At the FCC, the cross sections are two orders of magnitude larger than at the LHC, reaching $\rm \sigma(\gaga\to H)$~=~1.75~nb 
and 1.5~pb in Pb-Pb and p-Pb collisions at $\sqrtsnn$~=~39 and 63~TeV which, 
for the nominal $\LumiInt$~=~110~nb$^{-1}$ and 29~pb$^{-1}$ per-month integrated luminosities, yield $\sim$200 
and 45 Higgs bosons (corresponding to 110 and 25 bosons in the $\bbbar$ decay mode), respectively. 

\section{Data analysis and Higgs boson significances}

The observation of the Higgs boson in UPCs relies on the measurement of two exclusive b-jets with invariant masses peaked
at $m_{\rm H}$, on top of a background of exclusive $\gamma\gamma \rightarrow \bbbar, \ccbar, \qqbar$ continuum pairs, where charm 
and light (q = u,\,d,\,s) quarks are misidentified as b-quarks. For all colliding systems, 
the pure MC-level background continuum cross sections over $m_{\rm H}\approx 100$--150~GeV, computed with the same \madgraph~5 
setup described above, are about 25, 200, and 500 times larger respectively than the Higgs signal. Experimentally, triggering the online selection of 
such type of events is straightforward given their unique signature characterized by two back-to-back high transverse momentum 
($p_{T}$) jets in an otherwise empty detector. The data analysis follows the strategy first proposed in Ref.~\cite{dEnterria:2009cwl}, 
where more details, not repeated here, can be found. The following acceptance and reconstruction performances have been 
assumed: jet reconstruction over $|\eta|<2$ ($<5$ for FCC), 7\% b-jet energy resolution (resulting in a dijet mass resolution 
of $\sigma_{jj}\approx$\,6~GeV at the Higgs peak), 70\% b-jet tagging efficiency, and 5\% (1.5\%) b-jet mistagging probability 
for a c (light-flavour q) quark. We note that the b-jet reconstruction and identification performances are expected to be better 
in the very ``clean'' exclusive environment of UPCs than in the current high-pileup p-p collisions at the LHC~\cite{Sirunyan:2017ezt}. 
The pseudorapidity acceptance cut $|\eta|<2$ used for HL-LHC and HE-LHC keeps a large fraction of signal jets (around 80\%), while removing 
2/3 of the background jets, which are much more forward/backward-peaked. In terms of jet (mis)tagging efficiencies, for the double b-jet 
final-state of interest, these lead to a $\sim$50\% efficiency for the MC-generated signal ($\mathcal{S}$), and a total reduction 
of the misidentified $\ccbar$ and $\qqbar$ continuum backgrounds ($\mathcal{B}$) by factors of $\sim$400 and $\sim$4500 respectively. 
The sum of remaining continuum backgrounds can be further reduced through proper kinematical cuts by requiring~\cite{dEnterria:2009cwl}:
\begin{enumerate}[label=(\roman*),topsep=0pt, partopsep=0pt]
\setlength{\itemsep}{0pt}
\setlength{\parskip}{0pt}
\item both jets to have transverse momenta around half the Higgs mass, $p_{T} \approx m_{\rm H}/2\approx$~55--67~GeV, 
as expected for two back-to-back jets from the decay of an UPC Higgs produced almost at rest, thereby suppressing more than 95\% of the continuum, 
while removing about half of the signal; 
\item the angle of the jets to be within $|\cos \theta_{j_{1}j_{2}}| < 0.5$ 
--- to exploit the fact that the angular distribution in the helicity frame of the Higgs decay b-jets is isotropic 
while the continuum (with quarks propagating in the $t$- or $u$- channels) is peaked in the forward--backward directions --- 
further suppressing the backgrounds while leaving almost untouched the number of signal events; and 
\item the pair jet mass to be within $\pm1.4\sigma_{jj}$ around the Higgs mass (\ie\ $116 \lesssim m_{\bbbar} \lesssim 134$~GeV).
\end{enumerate}
For all systems, the overall loss of Higgs signal events due to the acceptance and kinematical cuts (\ie\ without accounting
for (mis)identification efficiencies) is around a factor of two, whereas the backgrounds are reduced by factors of 
30 to 100, resulting in a final $\mathcal{S}/\mathcal{B}\approx 1$ for all colliding species.\\

The current analysis is based on LO estimates for the signal and background cross sections. The contributions from
higher-order corrections to $\gaga\to$\,(H)\,$\to\bbbar$ and $\gaga\to\bbbar$, that can be significant in some regions of phase space, 
have been studied in detail in~\cite{NLO_bbbar} for the photon linear collider case. Our applied selection criteria, in particular 
the exclusive back-to-back 2-jets requirement and the $|\cos \theta_{j_{1}j_{2}}| < 0.5$ cut, effectively remove most of such higher-order contributions. 
Also, we note that any theoretical uncertainty on the photon fluxes impacts in a similar way the 
expected yields for signal and background, and thereby leaves the  $\mathcal{S}/\mathcal{B}$ ratio basically unaffected.
Last but not least, more advanced multivariate studies could be contemplated, rather than the simpler ``cut-based'' criteria applied here, 
that could further improve the separation of signal over background.\\

Table~\ref{tab:2} lists the cross sections after each event selection step, as well as the final number of events expected
(for the nominal integrated luminosities per run) for signal and backgrounds in the systems with larger signal 
strength at each collider (Fig.~\ref{fig:2}, right): Ar-Ar at $\sqrtsnn = 6.3$~TeV, Kr-Kr at $\sqrtsnn = 12.5$~TeV, 
and Pb-Pb at $\sqrtsnn = 39$~TeV, 
as well as the full MC results obtained for Pb-Pb and p-Pb at $\sqrtsnn = 39, 63$~TeV first discussed in~\cite{dEnterria:2017qte}.
The last column of Table~\ref{tab:2} lists the final number of signal and background events expected after all selection 
criteria for the nominal 1-month ($10^6$~s) run operation.
The expected number of Higgs per month, after cuts, at the HL-LHC and HE-LHC are below unity, whereas one expects
5 to 20 reconstructed H$(\bbbar)$ events at the FCC.

\begin{figure}[H]
\centering
\includegraphics[width=0.48\columnwidth]{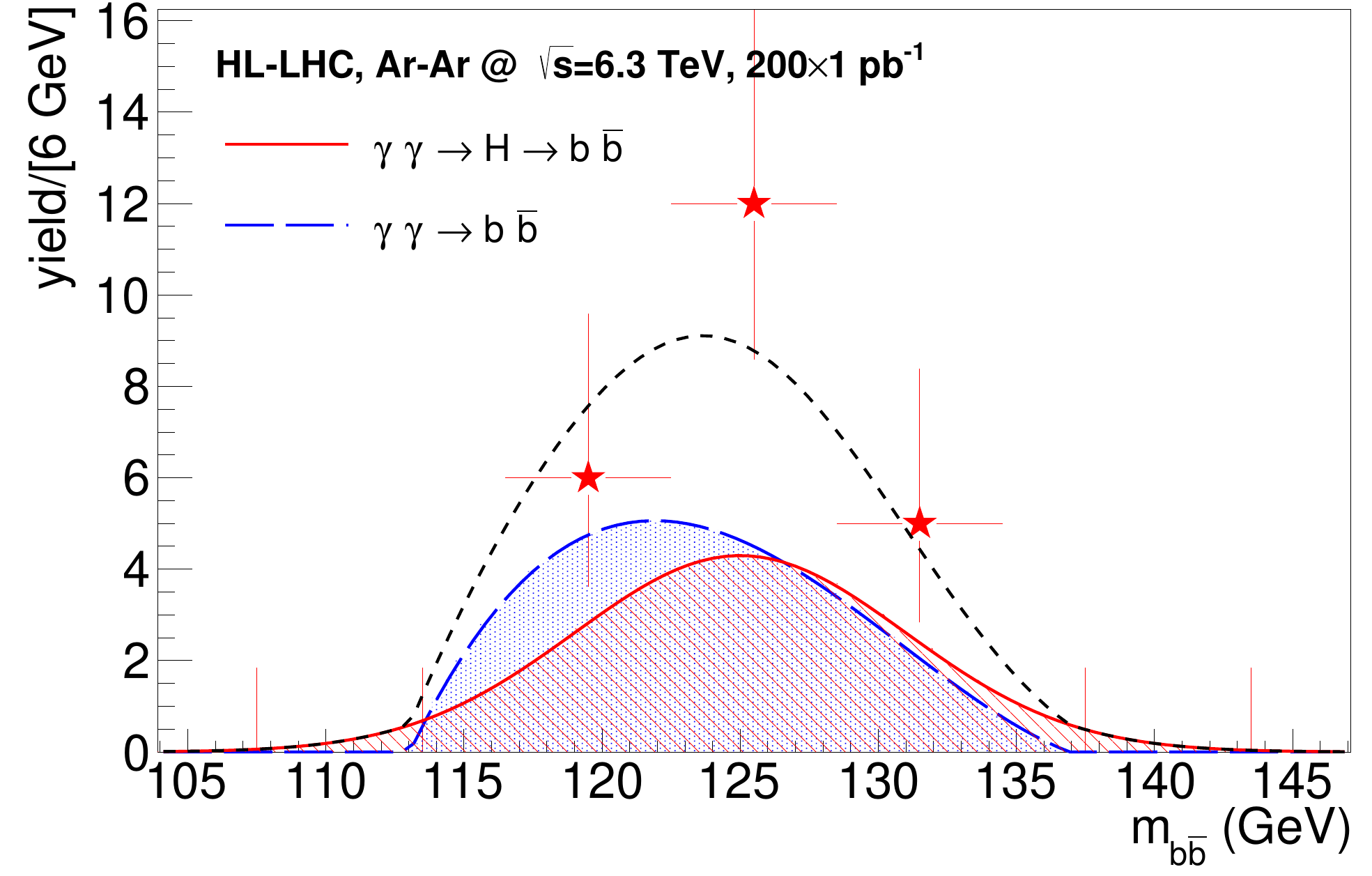}
\includegraphics[width=0.48\columnwidth]{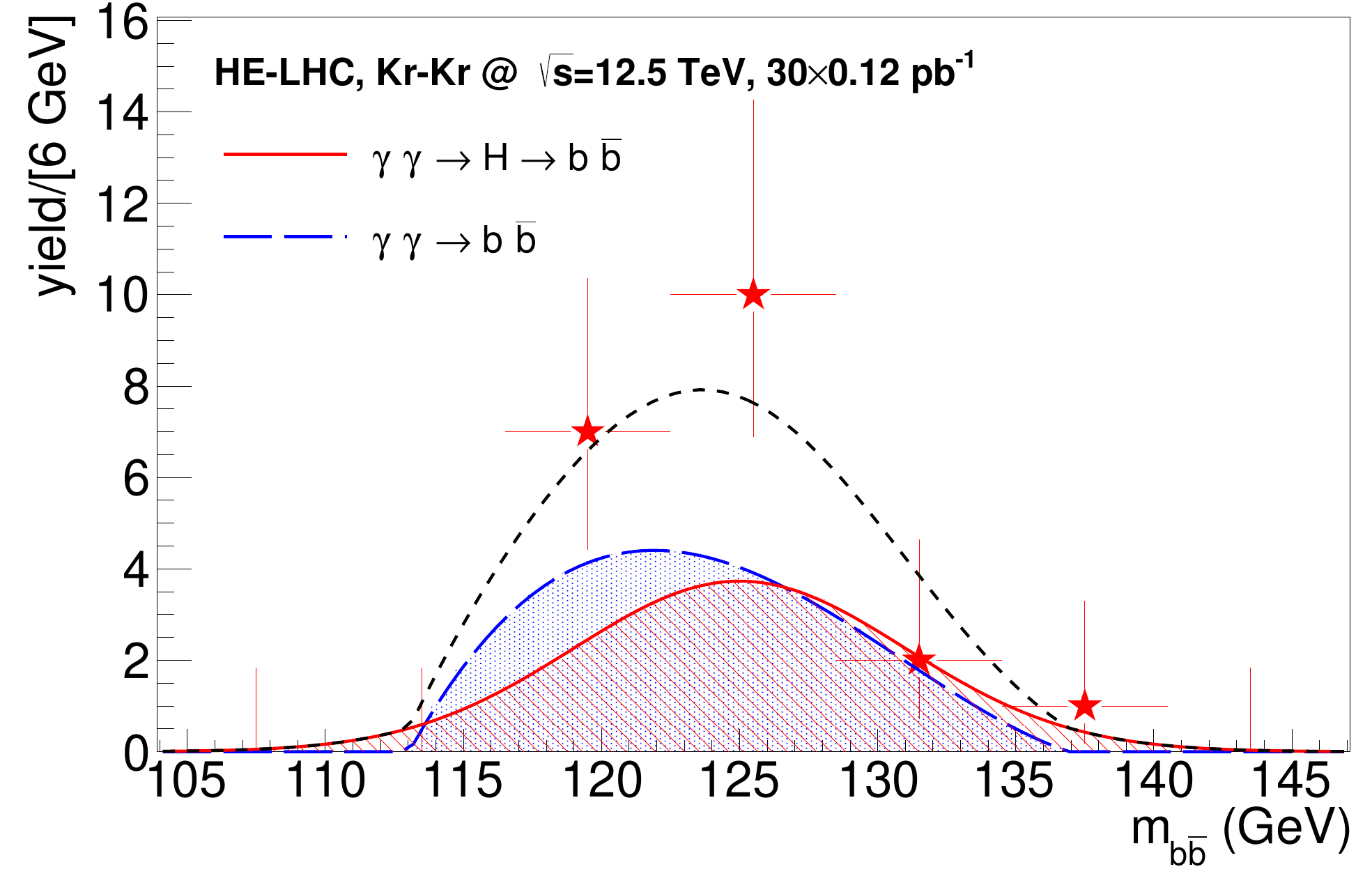}
\caption{\label{fig:3}Expected invariant mass distributions for b-jet pairs from the photon-fusion
Higgs signal (hatched red Gaussian) and $\bbbar+\ccbar+\qqbar$ continuum (hatched blue area) 
in ultraperipheral Ar-Ar ($\sqrtsnn = 6.3$~TeV, left) and Kr-Kr ($\sqrtsnn = 12.5$~TeV, right) 
collisions, after event selection criteria and with the quoted integrated luminosities (see text).
The red stars show the expected signal-plus-background invariant mass counts.
The dashed black curve corresponds to the sum of theoretical signal and background yields.}
\end{figure}

The final significance of the signal can be derived from the number of counts within $\pm1.4\sigma_{jj}$ around the Gaussian 
Higgs peak (\ie\ $116 \lesssim m_{\bbbar} \lesssim 134$~GeV) over the dijet continuum remaining after cuts.
In a simplified cut-and-count approach, one can estimate the statistical sample increase needed for a $3\sigma$ 
evidence from the $N_{\rm Higgs}$ values listed in the last column of Table~\ref{tab:2}. For an integrated luminosity 
two-hundred times larger than the nominal for Ar-Ar(6.3~TeV), one has $\mathcal{S}/\sqrt{\mathcal{B}}\approx 10/\sqrt{12}\approx 3$.
The same numbers at the HE-LHC, for thirty times more luminosity integrated in Kr-Kr(12.5~TeV), yield 
$\mathcal{S}/\sqrt{\mathcal{B}}\approx 9/\sqrt{10}\approx 3$. Thus, reaching 3$\sigma$ evidence of UPC Higgs-production 
at HL-LHC and at HE-LHC, requires at least factors of 
$\times$200 and $\times$30 more integrated luminosities in Ar-Ar and Kr-Kr collisions, respectively, than currently designed.
Figure~\ref{fig:3} shows the expected invariant dijet mass distributions after selection criteria for signal and 
backgrounds at the HL-LHC (Ar-Ar, left) and HE-LHC (Kr-Kr, right) for such increased integrated luminosities.
A factor of ten increase in $\LumiInt$ could be gained at both colliders simply by running during the time ($10^7$~s) 
typical of a proton-proton run, instead of the nominal 1-month( $10^6$~s) heavy-ion run operation. 
Such a longer run, motivated by Higgs- rather than heavy-ion physics, would allow for an evidence of the process 
at HE-LHC, by combining three experiments (or over three runs in a single one). Achieving the same significance 
at the HL-LHC seems out of reach, unless an extra factor of $\times$20 in the instantaneous Ar-Ar luminosity is accomplished 
by some currently unidentified means. In any case, going from the simple evidence to a $5\sigma$ observation would 
require yet another extra $(5/3)^2\approx 2.8$ increase in the collected data. These estimates indicate that the
UPC Higgs observation will very likely remain elusive at the HL-LHC and HE-LHC.\\


\begin{table}[H]
\centering
\caption{Summary of the cross sections after each event selection step (see text for details), and final number of events 
expected (for the nominal integrated luminosities quoted) for signal ($N_{\rm Higgs}$) and backgrounds ($N_{\rm backg}$) 
in the ${\rm \gaga \rightarrow H}(\bbbar)$ 
measurements in Ar-Ar at HL-LHC, Kr-Kr at HE-LHC, and Pb-Pb and p-Pb at FCC.}
\label{tab:2}
\vspace{0.2cm}
\begin{tabular}{lccc}\hline
Ar-Ar at $\sqrtsnn$ = 6.3 TeV             & cross section                      & visible cross section & $N_{\rm Higgs},\,N_{\rm backg}$  \\
             & (b-jet (mis)tag effic.)\;\; & \;\;after $\eta^j,p_T^j,|\cos\theta_{jj}|,m_{jj}$ cuts\;\; & ($\LumiInt$~=~1.1 pb$^{-1}$) \\\hline
$\gaga \rightarrow {\rm H} \rightarrow \bbbar$                              & 0.20~pb (0.10~pb) & 0.045 pb & 0.05 \\
$\gaga \rightarrow \bbbar$ \;$[\rm \scriptstyle{m_{\ensuremath\bbbar} = 100-150~GeV]}$ & 8.2~pb (4.0~pb) & 0.06 pb & 0.06 \\
$\gaga \rightarrow \ccbar$ \;$[\rm \scriptstyle{m_{\ccbar} = 100-150~GeV]}$ & 61~pb (0.15~pb) & 0.006 pb& 0.006 \\
$\gaga \rightarrow \qqbar$ \;$[\rm \scriptstyle{m_{\qqbar} = 100-150~GeV]}$ & 70~pb (0.016~pb)& $<10^{-3}$ & $<10^{-3}$ \\\hline

Kr-Kr at $\sqrtsnn$ = 12.5 TeV            &  &  & $N_{\rm Higgs}\,N_{\rm backg}$ \\
             &  &  & ($\LumiInt$~=~0.12 pb$^{-1}$) \\\hline
$\gaga \rightarrow {\rm H} \rightarrow \bbbar$                              & 11~pb (5.5~pb) & 2.5 pb & 0.30 \\
$\gaga \rightarrow \bbbar$ \;$[\rm \scriptstyle{m_{\ensuremath\bbbar} = 100-150~GeV]}$ & 365~pb (178~pb) & 2.8 pb & 0.34 \\
$\gaga \rightarrow \ccbar$ \;$[\rm \scriptstyle{m_{\ccbar} = 100-150~GeV]}$ & 2.7~nb (6.7~pb) & 0.24 pb& 0.03 \\
$\gaga \rightarrow \qqbar$ \;$[\rm \scriptstyle{m_{\qqbar} = 100-150~GeV]}$ & 3.1~nb (0.70~pb)& $<10^{-3}$ & $<10^{-4}$ \\\hline

Pb-Pb at $\sqrtsnn$ = 39 TeV            &  &  & $N_{\rm Higgs}\,N_{\rm backg}$ \\
            &  &  & ($\LumiInt$~=~110 nb$^{-1}$) \\\hline
$\gaga \rightarrow {\rm H} \rightarrow \bbbar$                              & 1.0~nb (0.50~nb) & 0.19 nb & 21.1 \\
$\gaga \rightarrow \bbbar$ \;$[\rm \scriptstyle{m_{\bbbar} = 100-150~GeV]}$ & 24.3~nb (11.9~nb)& 0.23 nb & 25.7 \\
$\gaga \rightarrow \ccbar$ \;$[\rm \scriptstyle{m_{\ccbar} = 100-150~GeV]}$ & 525~nb (1.31~nb) & 0.02 nb &  2.3 \\
$\gaga \rightarrow \qqbar$ \;$[\rm \scriptstyle{m_{\qqbar} = 100-150~GeV]}$ & 590~nb (0.13~nb) & 0.002 nb& 0.25 \\\hline

p-Pb at $\sqrtsnn$ = 63 TeV              &  &  & $N_{\rm Higgs}\,N_{\rm backg}$\\
             &  &  &  ($\LumiInt$~=~29 pb$^{-1}$) \\\hline
$\gaga \rightarrow {\rm H} \rightarrow \bbbar$                              & 0.87~pb (0.42~pb) & 0.16 pb & 4.8 \\
$\gaga \rightarrow \bbbar$  \;$[\rm \scriptstyle{m_{\ensuremath\bbbar} = 100-150~GeV]}$ & 21.8~pb (10.7~pb) & 0.22 pb & 6.3 \\
$\gaga \rightarrow \ccbar$ \;$[\rm \scriptstyle{m_{\ccbar} = 100-150~GeV]}$ & 410~pb (1.03~pb) & 0.011 pb& 0.3 \\
$\gaga \rightarrow \qqbar$ \;$[\rm \scriptstyle{m_{\qqbar} = 100-150~GeV]}$ & 510~pb (0.114~pb)& 0.001 pb& 0.04 \\\hline
\end{tabular}
\end{table}

\begin{figure}[H]
\centering
\includegraphics[width=0.48\columnwidth]{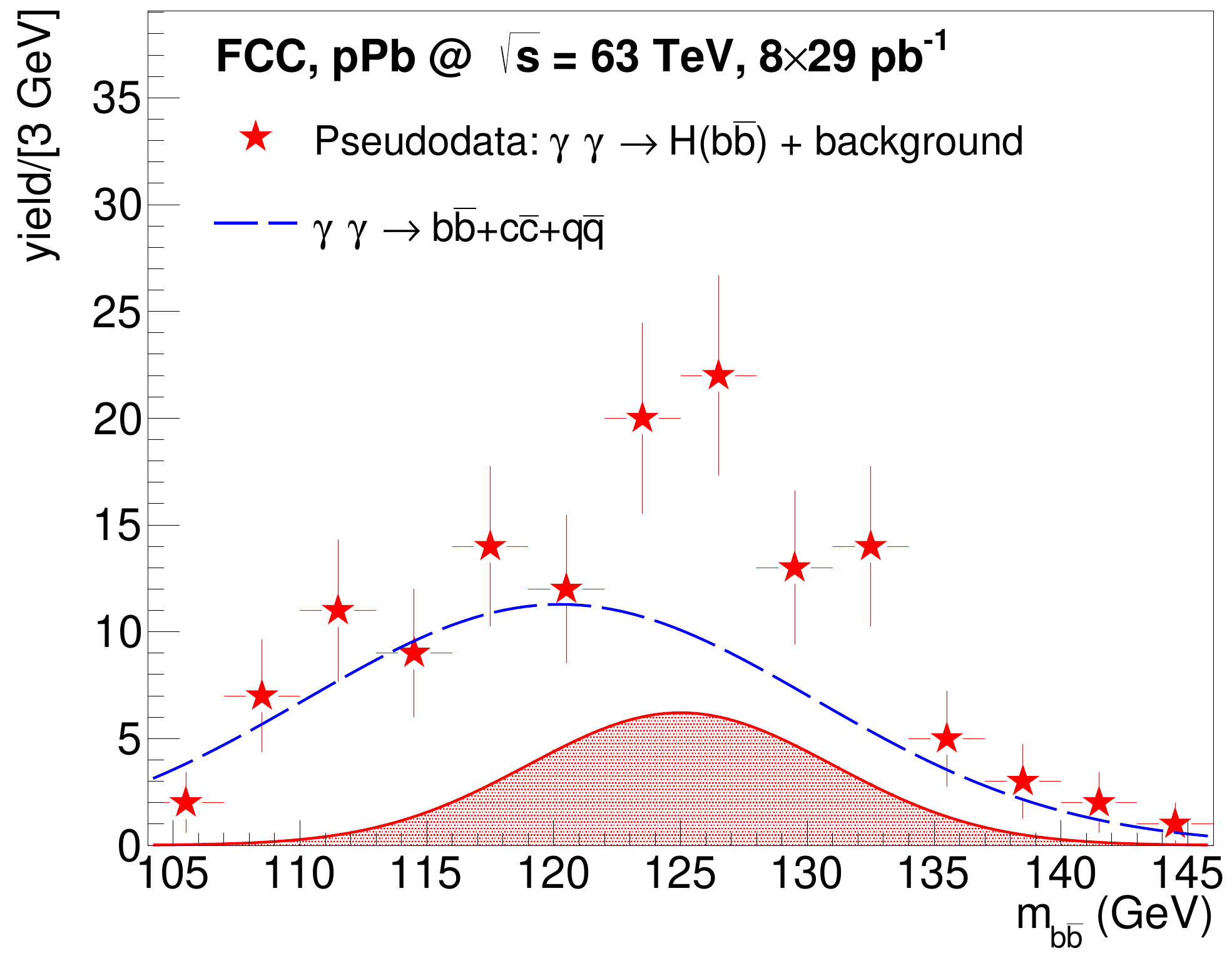}
\includegraphics[width=0.48\columnwidth]{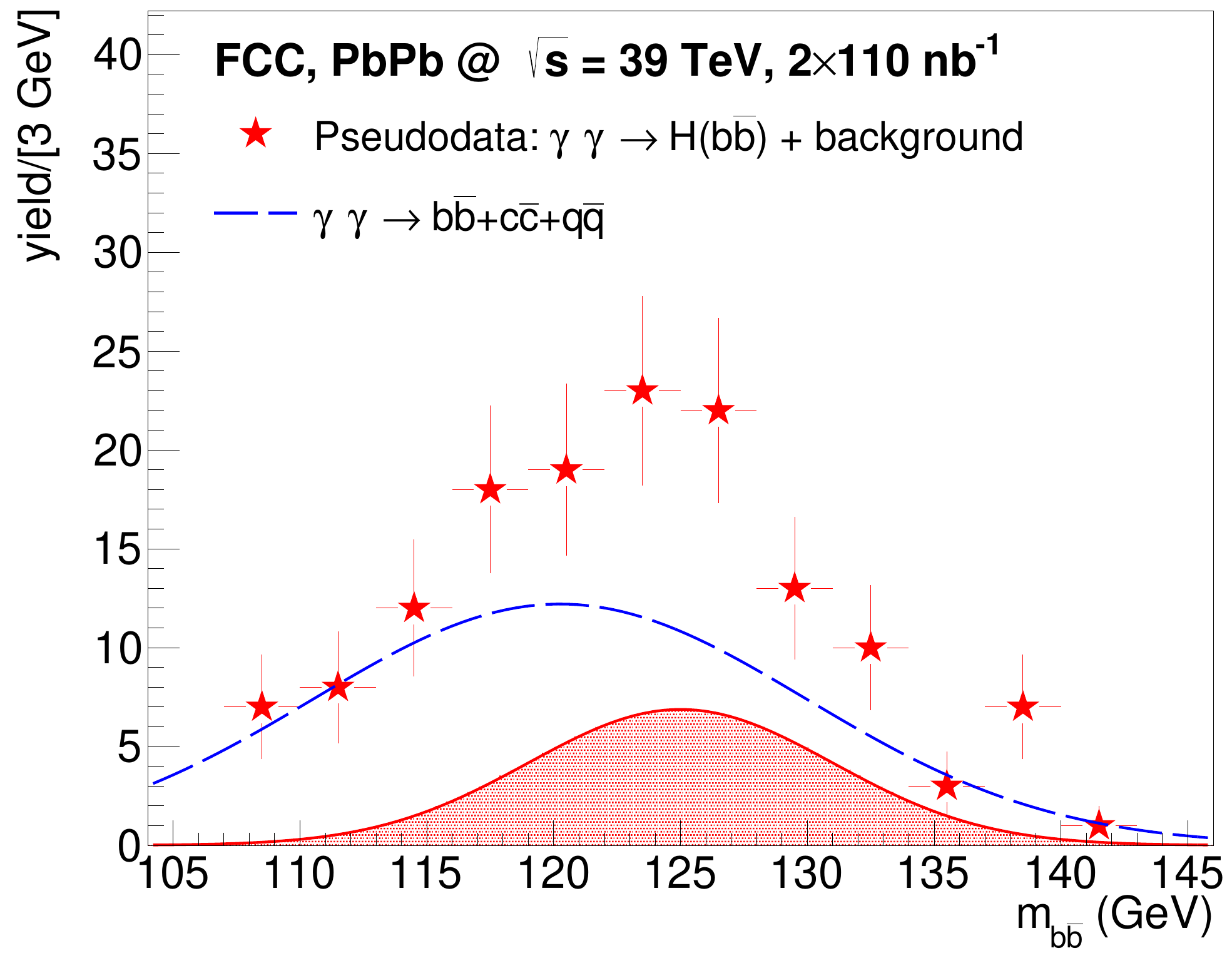}
\caption{\label{fig:4}Expected invariant mass distributions for b-jet pairs from the photon-fusion
Higgs signal (hatched red Gaussian) and $\bbbar+\ccbar+\qqbar$ continuum (blue curve) in ultraperipheral 
p-Pb ($\sqrtsnn = 63$~TeV, left) and Pb-Pb ($\sqrtsnn = 39$~TeV, right) collisions, after 
event selection criteria and with the quoted integrated luminosities (see text).
The red stars show the expected signal-plus-background invariant mass counts.}
\end{figure}

The situation appears much more favorable at the FCC thanks to the factors ten and one-hundred larger Higgs cross sections,
and factors ten increased instantaneous luminosities, compared to the HE-LHC and HL-LHC. Figure~\ref{fig:4} presents 
the expected double b-jet invariant mass distributions in p-Pb (left) and Pb-Pb (right) at the FCC. 
Lead-lead collisions at $\sqrts$~=~39 GeV with the nominal integrated luminosity of 
$\LumiInt = 110\;\mbox{nb}^{-1}$ per month, yield $\sim$20 signal counts over about the same number for the sum 
of backgrounds in a $m_{\bbbar} \approx 116$--134~GeV window. Reaching a $5\sigma$ statistical significance would just require 
to combine the measurements of the first run from two different experiments or accumulating two 1-month runs in a single one 
(Fig.~\ref{fig:4}, right). Similar estimates for p-Pb at 63~TeV yield about 5 signal events after cuts over a background
of 7 continuum events for the design $\LumiInt = 29$~pb$^{-1}$. Reaching a $5\sigma$ observation 
of $\gaga\to$~H production requires in this case to run for $\sim$8 months ($10^7$~s), instead of the 
nominal 1-month, or running 4 months and combining two experiments (Fig.~\ref{fig:4}, left).

All the derived number  of events and significances are based on the aforementioned simple set of kinematical cuts and 
signal-over-background estimates, and can be likely further improved by using more advanced multivariate 
studies and a full parametric shape analysis for the significance calculation. Notwithstanding such potential improvements, 
the numbers presented here provide realistic estimates of the feasibility of the UPC Higgs boson measurements 
at all currently (planned or under consideration) future CERN hadron colliders.

\section{Summary}

We have presented prospect studies for the measurement of the two-photon production of the Higgs boson 
in ultraperipheral Pb-Pb, Xe-Xe, Kr-Kr, Ar-Ar, O-O, p-Pb, and p-p collisions at three planned CERN future hadron colliders: 
HL-LHC, HE-LHC, and FCC. Cross sections have been obtained with \madgraph~5, modified to include the corresponding nuclear 
equivalent photon fluxes with no hadronic overlap of the colliding beams, for nucleon-nucleon \cm\ energies over 
$\sqrtsnn = 3$--100~TeV. The Higgs cross sections roughly rise by a factor of ten (one hundred) 
when increasing the \cm\ energy from the HL-LHC to the HE-LHC (FCC). At the HL-LHC and HE-LHC, although Pb-Pb features the 
largest Higgs cross section, $\rm \sigma(\gaga\to H)$~=~15, and 150~pb thanks to its $Z^4$-amplified photon fluxes, 
the most competitive systems to try a measurement of UPC Higgs production are Ar-Ar and Kr-Kr respectively, 
thanks to the larger available beam luminosities for such lighter species.\\

The observation of the Higgs boson in UPCs, via its dominant $\bbbar$ decay channel, relies on the measurement of 
two exclusive b-jets with invariant masses peaked at $m_{\rm H}$, on top of a background of 
$\gamma\gamma \rightarrow \bbbar, \ccbar, \qqbar$ continuum pairs, where charm and light (q = u,\,d,\,s) quarks 
are misidentified as b-quarks. The same \madgraph~5 setup used to compute Higgs cross sections and generate
the corresponding events has been employed for the exclusive two-photon production of $\bbbar$, $\ccbar$, and $\qqbar$ dijets. 
The HL-LHC and HE-LHC analyses have been carried out at the parton level, whereas for FCC energies the b-quarks 
have been showered and hadronized with \pythia~8, and reconstructed in a exclusive two-jet final-state with the 
$k_T$ algorithm. Given the simplicity of the exclusive final states considered, no significant differences 
between hadron- and parton-level results exist.
By assuming realistic jet acceptance, reconstruction performances, and (mis)tagging efficiencies, 
and applying appropriate kinematical cuts on the jet $p_T$ and angles, it has been shown that the H$(\bbbar)$ signal 
can be reconstructed on top of the $\gaga\to\bbbar, \ccbar, \qqbar$ continuum backgrounds. On the one hand, reaching 3$\sigma$ evidence 
of UPC Higgs-production at HL-LHC and at HE-LHC, requires factors of about $\times$200 and $\times$30 more 
integrated luminosities in Ar-Ar and Kr-Kr collisions, respectively, than currently planned for both machines. 
Factors of ten in integrated luminosity can be gained running for the duration ($10^7$ s) typical of a proton-proton run, 
rather than the nominal 1-month heavy-ion operation.
This would open up the possibility of a $3\sigma$ evidence at the HE-LHC, but would still fall too short for any feasible
measurement at the HL-LHC. On the other hand, the measurement of ${\rm \gaga \rightarrow H} \rightarrow \bbbar$ 
would yield about 20 (5) signal counts after cuts in Pb-Pb (p-Pb) collisions for their nominal integrated luminosities per run. 
Observation of the photon-fusion Higgs production at the $5\sigma$-level is achievable in the first FCC run by 
combining the measurements of two experiments (or doubling the luminosity in a single one) in Pb-Pb, 
and by running for about 8 months (or 4 months and combining two experiments) in the p-Pb case.
The feasibility studies presented here indicate the Higgs physics potential open to study in $\gamma\gamma$ ultraperipheral 
ion collisions at current and future CERN hadron colliders, eventually providing an independent measurement of the 
H-$\gamma$ coupling not based on Higgs decays but on a $s$-channel production mode, as well as of its total 
width combining the photon-fusion measurement with the H\,$\to\gaga$ decay branching ratio accessible at a
future $\epem$ collider.\\

\noindent {\bf Acknowledgments --} P.\,R.\,T. acknowledges financial support from the CERN TH Department and from the FCC project.
We thank I.~Helenius and L.~Harland-Lang for useful discussions on \pythia~8 and/or photon-photon collisions, as well
as J.~Jowett for feedback on running conditions for light-ions at the HL-LHC and HE-LHC, and E.~Chapon on statistical methods.

%

\end{document}